\documentclass{emulateapj}
\RequirePackage{lineno}
\usepackage{amssymb}
\usepackage{amsmath}
\usepackage{mathrsfs}
\usepackage{natbib}
\usepackage[colorlinks, linkcolor=blue, urlcolor=blue, citecolor=blue]{hyperref}
\usepackage{array}
\usepackage{float}
\usepackage{graphicx}
\usepackage{subfigure}
\usepackage{color}
\usepackage[all]{hypcap}
\usepackage[toc,page]{appendix}
\usepackage{multirow}
\usepackage{stfloats}
\usepackage{lineno}

\def\beq{\begin{equation}}
	\def\enq{\end{equation}}
\def\bea{\begin{eqnarray}}
	\def\ena{\end{eqnarray}}

\begin{document}

\title{The Propagation of Fast Radio Bursts in the Magnetosphere Shapes Their Waiting-time and Flux Distributions}
\author{Di Xiao\altaffilmark{1}, Zi-Gao Dai\altaffilmark{2}, Xue-Feng Wu\altaffilmark{1}}
\affil{\altaffilmark{1}Purple Mountain Observatory, Chinese Academy of Sciences, Nanjing 210023, People's Republic of China; dxiao@pmo.ac.cn}
\affil{\altaffilmark{2}Department of Astronomy, University of Science and Technology of China, Hefei 230026, China}

\begin{abstract}
The field of fast radio bursts (FRBs) has entered the age of fine characterization as observational results from different radio telescopes become more and more abundant. The large FRB sample is suitable for a statistical study. There is an interesting finding that the waiting-time distributions of very active repeating FRBs show a universal double-peaked feature, with left peaks lower than right ones. Assuming these two peaks are independent and initially comparable, we show that the observed asymmetric shape can be ascribed to the propagational effect in the magnetosphere. An FRB passing through the magnetized plasma will induce the circular motion of charged particles to form a current loop. This further leads to an induced magnetic field with opposite direction respect to the background field. As the effective field strength changes, the scattering absorption probability of the following FRB will be influenced. The absorption can be important under certain physical conditions and bursts with smaller time-lags are easier to be absorbed. Also there will be an imprint on the flux distribution as the scattering optical depth depends on burst luminosity as well. 
\end{abstract}


\section{Introduction}
Fast radio bursts (FRBs) are very bright radio pulses with millisecond durations. Since the first discovery by Parkes telescope in 2007 \citep{Lorimer2007}, the total number of FRBs accumulate gradually, refreshing our understanding on this phenomenon continuously \citep[for reviews, see e.g.,][]{Cordes2019,Petroff2021,XiaoD2021,Zhang2022b}. Especially in recent years, there are growing interest in FRBs among the community and more and more radio facilities are involved in FRB observation. Canadian Hydrogen Intensity Mapping Experiment (CHIME) has a big advantage in finding new FRBs due to very large field of view and their first catalog has been released \citep{CHIME2021cat}. The interferometric arrays like Australian Square Kilometre Array Pathfinder (ASKAP) and Deep Synoptic Array (DSA) can localize FRBs to arcsecond precision \citep{Bannister2019, Law2023}. The most sensitive single-dish telescope at present, i.e., the Five-hundred-meter Aperture Spherical Telescope (FAST) is extremely powerful in the fine characterization of repeating FRBs. A lot of interesting properties of FRBs has been firstly found by FAST such as polarization angle swing \citep{Luo2020b}, bimodal energy distribution \citep{LiD2021}, dynamic magnetized environment \citep{XuH2022} and so on.

While the observation is making big progress, the physical mechanism of FRBs is still largely unknown. The association of FRB 20200428A with the X-ray bursts of SGR 1935+2154 confirmed that magnetars can produce FRBs \citep{CHIME2020a,Bochenek2020}. Under the scenario of a magnetar source, basically there are two major kinds of radiation models. One kind assumes FRBs are produced via synchrotron maser mechanism of relativistic shocks outside the magnetosphere \citep[e.g.,][]{Metzger2019,Beloborodov2020}. The other kind deems the production site is inside the magnetosphere via a pulsar-like mechanism \citep[e.g.,][]{Lu2018, Yang2018, Lyubarsky2020, Lyutikov2021a, Zhang2022}. These two kinds of models have different predictions that matched the FRB observation to a certain extent \citep{Margalit2020,XiaoD2020,Lu2020b,Yang2020b}, however, they certainly need to be tested by more observational results. 

The large sample of bursts by FAST make it possible for us to study the statistical properties of individual FRBs in detail. The bimodal energy distribution of FRB 20121102A motivates the discussion on FRB classification method \citep{XiaoD2022a}, further the constraints on FRB radiation mechanism \citep{XiaoD2022b}. However, the bimodal waiting-time distribution of FRB 20121102A is rarely discussed, as also found by Arecibo telescope later \citep{Hewitt2022}. More intriguingly, several other active repeaters show similar double-peaked waiting-time distributions, with left peaks being always lower than the right ones \citep{XuH2022,ZhangYK2022,ZhangYK2023}. This motivates us to study the physical reasons for the waiting-time distribution. Here in this work we believe it is caused by the propagational effect in the magnetosphere. 

This paper is organized as follows. We study the effect of wave-particle interaction in a magnetized environment in Section \ref{sec2}. An induced magnetic field is formed by the gyrating electrons under ponderomotive force. Then in Section \ref{sec3} we discussed the impact of this time-evolving induced field on observed FRB distributions. In Section \ref{sec4} we compare our theoretical predictions with observational results. We finish with discussion and conclusions in Section \ref{sec5}.

\section{Wave-particle Interaction and the effect of Ponderomotive Force}\label{sec2}
FRBs are strong waves and will accelerate the nearby charged particles since the dimensionless parameter 
\bea
a_0&=&\frac{eE}{m_ec\omega}\simeq\frac{e}{2\pi m_ec\nu}\left(\frac{4\pi\nu F_\nu}{c}\right)^{1/2}\nonumber\\
&=&1.77\times10^5\left(\frac{F_\nu}{\rm Jy}\right)^{1/2}\left(\frac{\nu}{\rm GHz}\right)^{-1/2}\left(\frac{D_{\rm L}}{\rm Gpc}\right)r_8^{-1}
\label{eq:a0}
\ena
is much larger than unity within 1 AU \citep{Yang2020a}, where $\omega$ is FRB circular frequency and $F_\nu$ is flux density. $D_{\rm L}$ is the luminosity distance and $r$ is the radial distance from FRB source. Since FRBs are not perfectly monochromatic plane waves, the electric field probably contains the spatial dependence ${\bf E=E}({\bf r})\exp{(- i\omega t)}$. Therefore the particles will feel a non-linear force called the ponderomotive force ${\bf F}_{\rm p}=-\frac{1}{4}\frac{e^2}{m_e\omega^2}\nabla E^2$ and follow the ``figure of eight" trajectory in an unmagnetized plasma \citep{Yang2020a}.

Consider a typical magnetar with spin period $P\sim 1\,\rm s$ and surface magnetic field $B_{\rm s}\sim 10^{15}\,\rm G$, the light cylinder radius is 
\beq
R_{\rm LC}=\frac{c}{\Omega}=\frac{cP}{2\pi}=4.78\times10^9 P {\,\rm cm},
\enq
and the Goldreich-Julian number density is \citep{Goldreich1969}
\bea
n_{\rm GJ}=\Omega B/(2\pi e c)=6.94\times10^7B_{{\rm s},15}P^{-1}r_8^{-3}\,{\rm cm^{-3}},
\label{eq:nGJ}
\ena
where we assume dipole field $B(r)=B_{\rm s}(r/R_{\rm NS})^{-3}$ and neutron star radius $R_{\rm NS}\sim 10^6\,\rm cm$. The cyclotron and plasma frequencies are  
\bea
\omega_B&=&\frac{eB}{m_ec}=1.76\times10^{16}B_{{\rm s},15}r_8^{-3}{\,\rm rad\,s^{-1}}\\
\omega_{\rm p}&=&\left(\frac{4\pi n_ee^2}{m_e}\right)^{1/2}\nonumber\\
&=&1.49\times10^{10}\mathcal{M}_3^{1/2}B_{{\rm s},15}^{1/2}P^{-1/2}r_8^{-3/2} {\,\rm rad\,s^{-1}},
\ena
where $n_e=\mathcal{M}n_{\rm GJ}$ and $\mathcal{M}$ is the multiplicity. Therefore in the magnetosphere  $\omega_B\gg\omega,\,\omega_{\rm p}$ is always satisfied. If a strong external magnetic field exists in the plasma, the motion of electrons is very different from the unmagnetized case. Instead of oscillations with momentum $\sim a_0m_ec$, electrons will experience $\bf E \times B$ drift and the ponderomotive force becomes ${\bf F}_{\rm p}^{(B)}=-\frac{m_ec^2}{4B_0^2}\nabla E^2$ \citep{Lyutikov2020b}.  The drift is in the azimuthal direction with a velocity\beq
{\bf v}_{\rm d}=\frac{c}{e}\frac{{\bf F}_{\rm p}^{(B)}\times{\bf B} }{B ^2}.
\enq
This drift motion of charged particles leads to the formation of a current loop, which will induce an opposite magnetic field with strength \citep{Lyutikov2020b} 
\beq
B_{\rm ind}= \pi n_{e,\rm b} m_ec^2\frac{E^2}{B ^3}.
\label{eq:Bind}
\enq
Here we introduce a ``background" non-relativistic plasma number density $n_{e,\rm b}=f n_e$ at each radius. Generally the electron-positron pairs are accelerated to relativistic speeds by parallel electric field at the beginning. As the electrons and positrons move in opposite directions along the magnetic field lines, two stream instability may develop and generate plasma oscillations that broaden the momentum distribution. While a large portion of pairs still flow outwards with high Lorentz factors and form a steady wind, a notable fraction of pairs are non-relativistic, as indicated by numerical simulation results \citep{Beloborodov2013a}. These slow-moving particles constitute a background plasma, however, the fraction $f$ is highly uncertain and depends on initial conditions. Basically the region of influence by the outgoing pairs would likely move upwards and quickly escape from the magnetosphere, leaving negligible impact on the following bursts. Therefore we only consider the induced field of non-relativistic particles. The induced current could only last for finite time because the non-relativistic plasma is not an ideal MHD fluid and electrical resistance exists. The Spitzer resistivity of the plasma is
\bea
\eta=\frac{\pi e^2m_e^{1/2}\ln\Lambda}{(k_BT_e)^{3/2}},
\ena
where the Coulomb logarithm is
\bea
\ln\Lambda=\ln\left[\frac{3k_B^{3/2}}{2\pi^{1/2}e^3}\left(\frac{T_e^3}{n_{e,\rm b}}\right)^{1/2}\right],
\ena
and $T_e$ is the plasma temperature. Generally, thermal equilibrium is not reached in the magnetosphere and the radial variation of $T_e$ is poorly understood. The background electrons are non-relativistic therefore the upper limit is $T_e\lesssim m_ec^2/k_B\sim10^{10}\,\rm K$. Physically, we may define an “effective temperature” according to the energy of charged particles. \citet{Lu2020c} suggested that due to the induced relativistic motion by the nonlinear effect, the effective temperature would be very high ($\sim a_0^2m_ec^2/k_B$) and dependent of the burst intensity. However, in the presence of an external strong magnetic field, this effect is suppressed since the magnetic nonlinearity parameter is
\bea
a_0^{(B)}&=&\frac{eE}{m_ec\omega_B}=a_0\frac{\omega}{\omega_B}\nonumber\\
&=&0.063\left(\frac{F_\nu}{\rm Jy}\right)^{1/2}\left(\frac{\nu}{\rm GHz}\right)^{1/2}\left(\frac{D_{\rm L}}{\rm Gpc}\right)B_{\rm s,15}^{-1}r_8^{2},\quad
\label{eq:a0B}
\ena
then the effective temperature is approximated as $T_e\sim (a_0^{(B)})^2m_ec^2/k_B$. Therefore $T_e$ varies in the range of $10^4-10^{10}\,\rm K$ among different bursts. For practical reason we just assume several typical values of $T_e$ in our calculation. Consider a loop of current, energy conservation means the energy of drift motion is converted to Joule heat
\bea
\frac{d}{dt}(\frac{1}{2}m_ev_{\rm d}^2n_{e,\rm b})=j^2\eta t,
\ena
Substituting the electric current density $j=ev_{\rm d}n_{e,\rm b}$ and we obtain
\bea
v_{\rm d}(r,t)=v_{\rm d,0}e^{(-\eta e^2n_{e,\rm b}/m_e)t}. 
\ena
Since $B_{\rm ind}\propto j\propto v_{\rm d}$, the time evolution of $B_{\rm ind}$ follows the same exponential decay. Therefore, an FRB can alter the field strength on its path, then the next FRB wave will propagate in an environment with effective field strength $B_{\rm eff}=B-B_{\rm ind}$. The characteristic dissipation time of the current is 
\beq
t_{\rm diss}=m_e/(\eta e^2n_{e,\rm b}).
\label{eq:tdiss}
\enq
Once the FRB burst ends, the induced magnetic field would become zero due to $E=0$. Therefore, The relaxation time of magnetic field to the initial status $t_{\rm re}$ is the minimum between burst duration $t_{\rm dur}$ and dissipation time $t_{\rm diss}$. This implies that the induced field can exist no longer than typical FRB duration of a few milliseconds. Observationally, subburst structures or multi-peak bursts are very common for active repeaters and they have extremely short time-lags, for which the induced field may play an important role.

\section{Transparency of FRBs in the magneosphere with a time-evolving field strength}
\label{sec3}
\citet{Beloborodov2021} argued that an FRB wave must interact with the pair plasma in the outer magnetosphere and the scattering cross section is so huge that radio waves cannot escape. However, \citet{Qu2022} found that the scattering optical depth can be less than unity as long as the pairs are moving outwards with a ultrarelativisic speed of $\gamma_p\geq10^3$, and the angle between magnetic field with FRB wave vector is small in the open field line region. These two requirements are not always satisfied for all magnetars then the absorption may occur. Following the assumptions of \citet{Qu2022}, we re-calculate the optical depth for our dynamic magnetic field scenario below.

Considering a beamed FRB emission with opening angle $\theta_j$, and around the pole, the angle $\theta_B$ between a dipole field line and the radial direction is $\theta_B\sim\theta_j$. In the plasma co-moving frame, Dopper correction is $\theta_B^{\prime}={\rm acrsin}(\mathcal{D}\sin\theta_B)$, where $\mathcal{D}=1/(\gamma_p(1-\beta_p\cos\theta_B))$. Same as \citet{Qu2022} we can define a critical radius $R_\theta$ where in the comoving frame $\omega_B^{\prime}/\omega^{\prime}=a\theta_B^{\prime}$ is satisfied. Within $R_\theta$ we have $\omega_B^{\prime}/\omega^{\prime}>a\theta_B^{\prime}$ and the scattering cross section drops dramatically. Therefore, the total scattering optical depth is
\bea
\tau =\int_{R_{\min}}^{R_{\max}}{n_e \sigma^{\prime}\gamma_p^2\frac{(1-\beta_p\cos\theta_B)^3}{\cos\theta_B}dr},
\ena 
where the integration ranges are $R_{\min}=\max{(R_{\rm FRB}, R_\theta)}$ and $R_{\max}=R_{\rm LC}$. The scattering with outgoing pairs and background plasma can be equally important and we add up the optical depth contributed by both of them. Throughout this work we assume that FRBs are produced at a radius of $R_{\rm FRB}\sim100R_{\rm NS}=10^8\,\rm cm$. Adopting the numerical results of cross section $\sigma^{\prime}$ in \citet{Qu2022}, we can calculate $\tau$ for consecutive FRBs.

As we can see from Eq.(\ref{eq:Bind}), the induced field is stronger as long as the incident FRB flux is higher or the background field is weaker. This means that the effective field strength is weaker at each radius after brighter FRBs pass by. Therefore, whether an FRB can escape the magnetosphere depends not only on the luminosity itself, but also the time interval and luminosity of previous bursts. More specifically, $R_\theta$ moves inward for an incident FRB with a higher flux, leading to a larger optical depth and higher probability for the next FRB being chocked. However if the time-lag is long, the magnetic field can recover to the initial static status then the influence of previous bursts is negligible. 

The direct impact of this time-evolving magnetic field is that the observed waiting-time and luminosity distribution will be strongly modified. In the case of significant absorption, the observed arriving time of consecutive FRBs is no longer independent. Since more luminous burst leads to a larger $B_{\rm ind}$, the next burst, if also bright with a short time-lag, is very likely to be chocked. On the contrary, less-luminous bursts can pass freely. Observationally, we may expect that usually a bright pulse is followed by a few weak pulses, and the situation that two bright pulses occur closely is somewhat uncommon. This has been confirmed by the observations of FRB 20121102A, where the median waiting time of high-energy bursts are clearly longer than that of low-energy ones \citep{ZhangGQ2021,Hewitt2022}.

\begin{figure}
	\label{fig1}
	\begin{center}
		\includegraphics[width=0.45\textwidth]{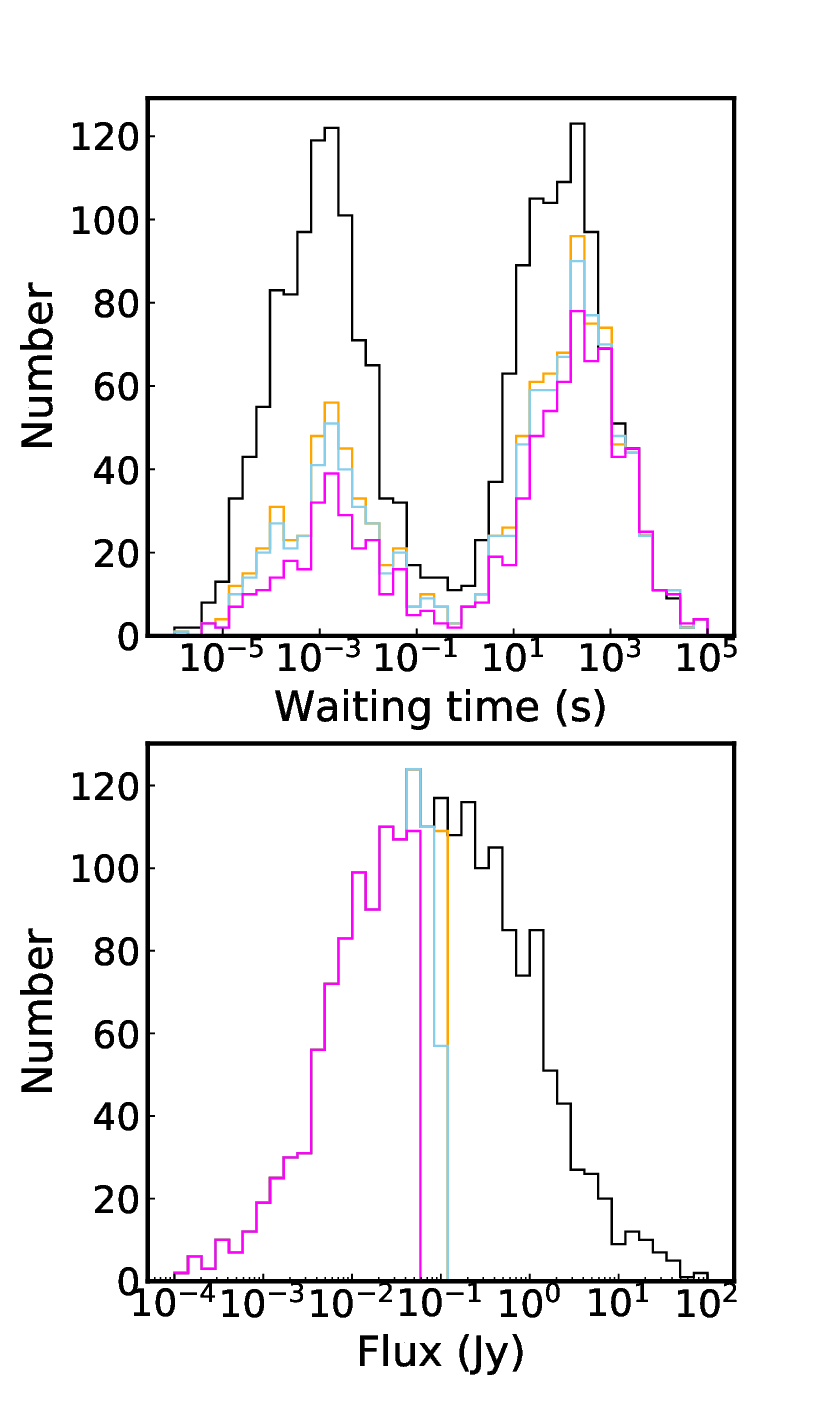}
		\caption{The waiting-time and flux distributions of mock FRB sample. Orange, sky blue and magenta colors represent for $\gamma_{\rm p}=100,\,30, \,10$ cases respectively, the distributions of which are clearly shaped by absorption. The initial distribution is shown in black color. We fix $\theta_B=0.2$ here.}
	\end{center}
\end{figure}

\begin{figure}
	\label{fig2}
	\begin{center}
		\includegraphics[width=0.48\textwidth]{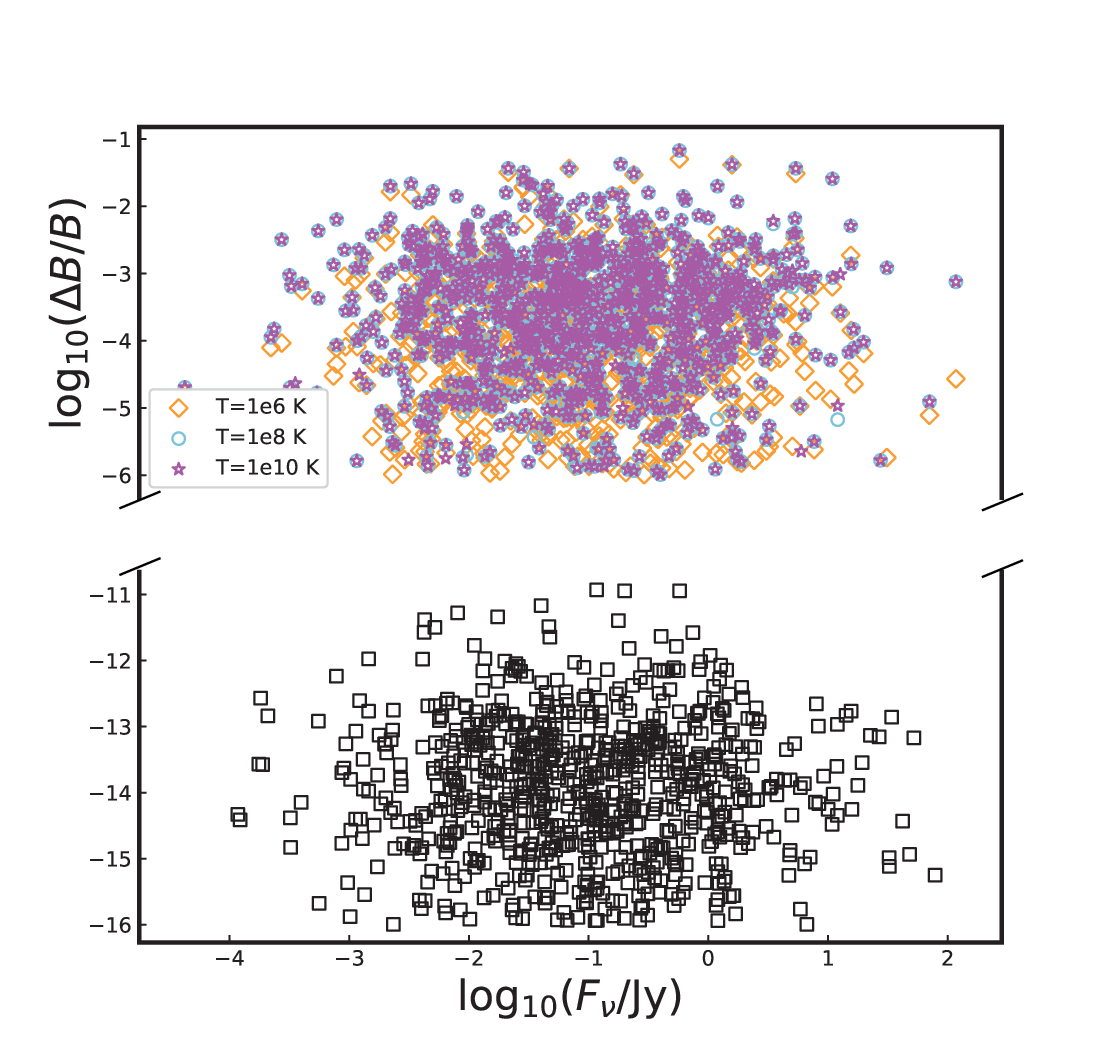}
		\caption{The relative strength of the induced magnetic field versus the flux of each burst in the sample. Using the parameters in Figure \ref{fig1}, black squares in the bottom shows that the induced field is negligible. However, significant absorption still occurs and shapes the distributions in Figure \ref{fig1} mainly due to relative low $\gamma_{\rm p}$ and large $\theta_B$. Adopting a different parameter set as in Figure \ref{fig3}, the three colored symbols in the top indicate that the induced field can be important and the effect of different plasma temperature becomes visible.}
	\end{center}
\end{figure}

\begin{figure}
\label{fig3}
\begin{center}
\includegraphics[width=0.45\textwidth]{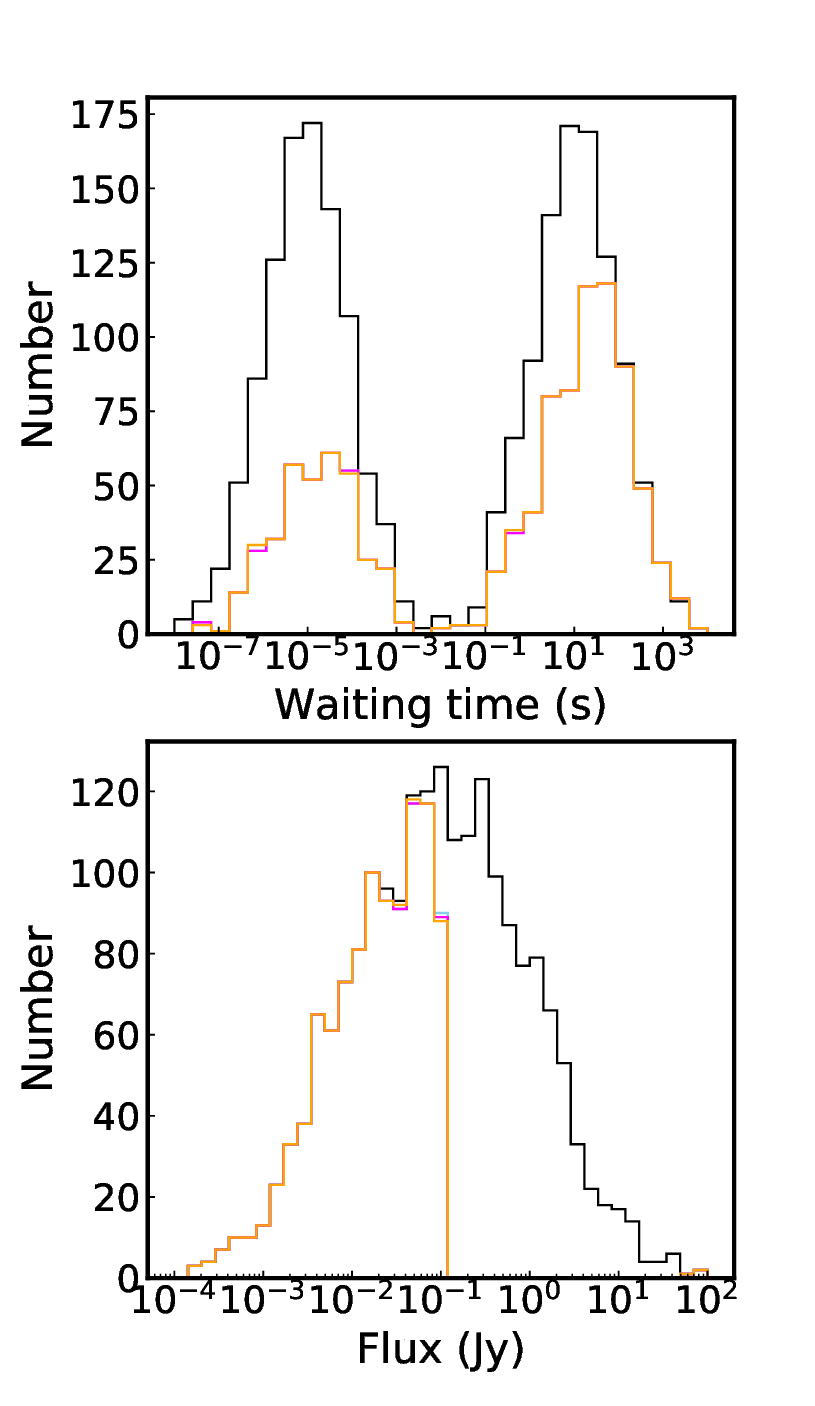}
\caption{Similar to Figure \ref{fig1} but using a different set of parameters with $B_{\rm s}=5\times10^{13}\,\rm G$, $P=0.1\,\rm s$, $\mathcal{M}=3\times10^6$, $f=0.2$, $\gamma_{\rm p}=1000$, $\theta_B=0.03$. Orange, sky blue and magenta colors represent $T_e=10^6,\,10^8,\,10^{10}\,\rm K$ respectively. The induced field becomes important as shown in Figure \ref{fig2}, leading to a distinguishable impact on the absorption probability. }
\end{center}
\end{figure}

\begin{figure}
	\label{fig4}
	\begin{center}
		\includegraphics[width=0.45\textwidth]{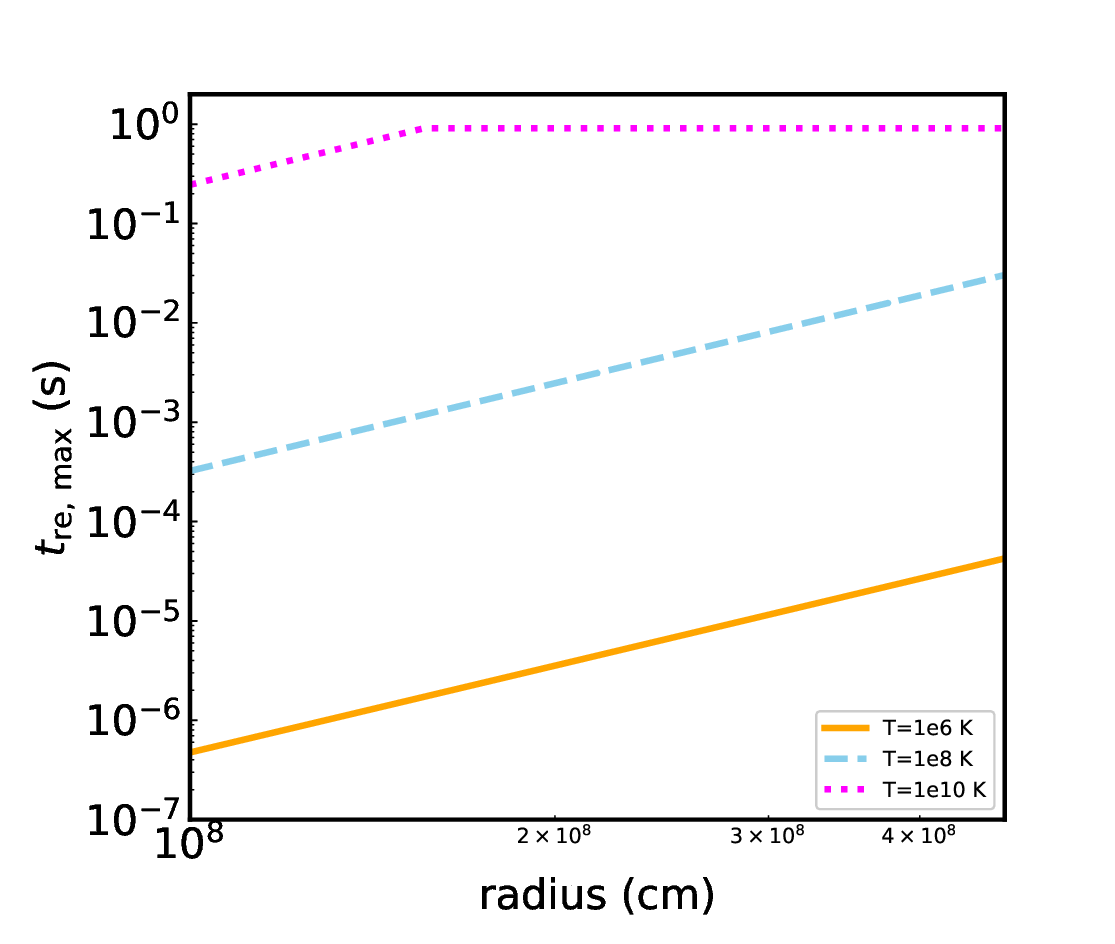}
		\caption{The corresponding relaxation time for the three cases in Figure \ref{fig3}. The Coulomb logarithm is not sensitive to $n_e$ in the radius range of interest. Therefore according to Eq.(\ref{eq:tdiss}), $t_{\rm re}$ is nearly in proportion to $r^3$. As the plasma temperature goes higher, the resistivity is smaller and it takes more time to dissipate the energy of gyration. For $T_e=10^{10}\,\rm K$ case, $t_{\rm re}$ at large radius flattens due to the limitation of maximum burst duration in the sample.}
	\end{center}
\end{figure}

To illustrate how much the degree of influence on the distributions could be, we mock a random sample of 2000 FRBs from one repeater. These FRBs has an intial log-normal flux distribution and bimodal log-normal waiting-time distribution. The centeral value and standard deviation of normal distributions are assumed somewhat arbitrarily. In Figure \ref{fig1} we show an example case in black color of $\log (F_{\nu,c}/{\rm Jy})=-1.0$, $\log(\Delta t_{1,c}/{\rm s})=-3$, $\log(\Delta t_{2,c}/{\rm s})=2$ with three standard deviations setted as $1.0$. The burst duration in the sample is also mocked as log-normal distribution of $\log(t_{\rm dur}/{\rm s})=-3$ with standard deviation 0.5. Further, we assume typical magnetar parameters $B_{\rm s}=10^{15}\,\rm G$, $P=1\,\rm s$, $\mathcal{M}=10^3$, $f=0.1$, $T_e=10^8\,\rm K$ and this repeater locates at a distance of 1 Gpc. We fix $\theta_B=0.2$ and plot three cases with different $\gamma_{\rm p}$. Orange, sky blue and magenta distributions represent cases of $\gamma_{\rm p}=100,\,30,\,10$ respectively. Bursts on the high flux end are absorbed as expected. More interestingly, the left peak of waiting-time distribution decrease dramatically. To illustrate the influence of the induced field, we plot in Figure \ref{fig2} the relative variation of magnetic field strength as a function of burst flux. Surprisingly, the black squares in the lower part of Figure \ref{fig2} indicates that the induced field is unimportant under this set of magnetar parameters, however, significant absorption still occur mainly due to relative low $\gamma_{\rm p}$ and large $\theta_B$ \citep{Qu2022}. The waiting time distribution becomes asymetric even though there is no preference of absorption on bursts with long or short time-lags. Nevertheless, for a different set of parameters the induced field becomes notable as shown in the upper part of Figure \ref{fig2},  further the influence on waiting time and flux distribution is illustrated in Figure \ref{fig3} correspondingly. Here we adjust two characteristic values of our mock sample as $\log(\Delta t_{1,c}/{\rm s})=-5$, $\log(t_{\rm dur}/{\rm s})=-2$ to create favorable conditions for absorption. Further we adopt a high multiplicity $\mathcal{M}=3\times10^6$ and fraction $f=0.2$ that leads to more abundant non-relativistic electrons and larger induced field according to Eq.(\ref{eq:Bind}). We note that these two values may be higher than normally assumed especially for $f$, because we expect most cascaded pairs have relativistic speeds. However, two stream instability may develop and significantly change the momentum distribution of pairs \citep[see the simulation result Figure 6 of][]{Beloborodov2013a}. Therefore, these two values are just achievable though not very typical, and we adopt them to make the effect of different $T_e$ visible. The plasma temperature plays an important role now because the relaxation time depends sensitively on it, which is shown in Figure \ref{fig4}. This leads to a marginally distinguishable impact on the observed distributions shown in Figure \ref{fig3}. Since the duration of each burst is different in our sample, here we plot the maximum relaxation time at different radius. For $T_e=10^{10} \,\rm K$ case, $t_{\rm diss}$ is large then $t_{\rm re}$ is limited by burst duration at large radius, while for the other two cases $t_{\rm re}$ is dominated by $t_{\rm diss}$. Once the plasma electrons do not have enough time to stop gyration induced by the former burst, smaller $B_{\rm eff}$ leads to larger possibily of the latter burst being chocked. For longer waiting time on the right peak, this induced magnetic field is unimportant mainly because the burst duration can not be that long. Therefore in the situation of significant absorption, the left peak should be always lower than the right one even if initially they are almost comparable, and this has been confirmed by the observations of several active repeaters, which will be discussed in the next section. 

\begin{figure}
	\label{fig5}
	\begin{center}
		\includegraphics[width=0.45\textwidth]{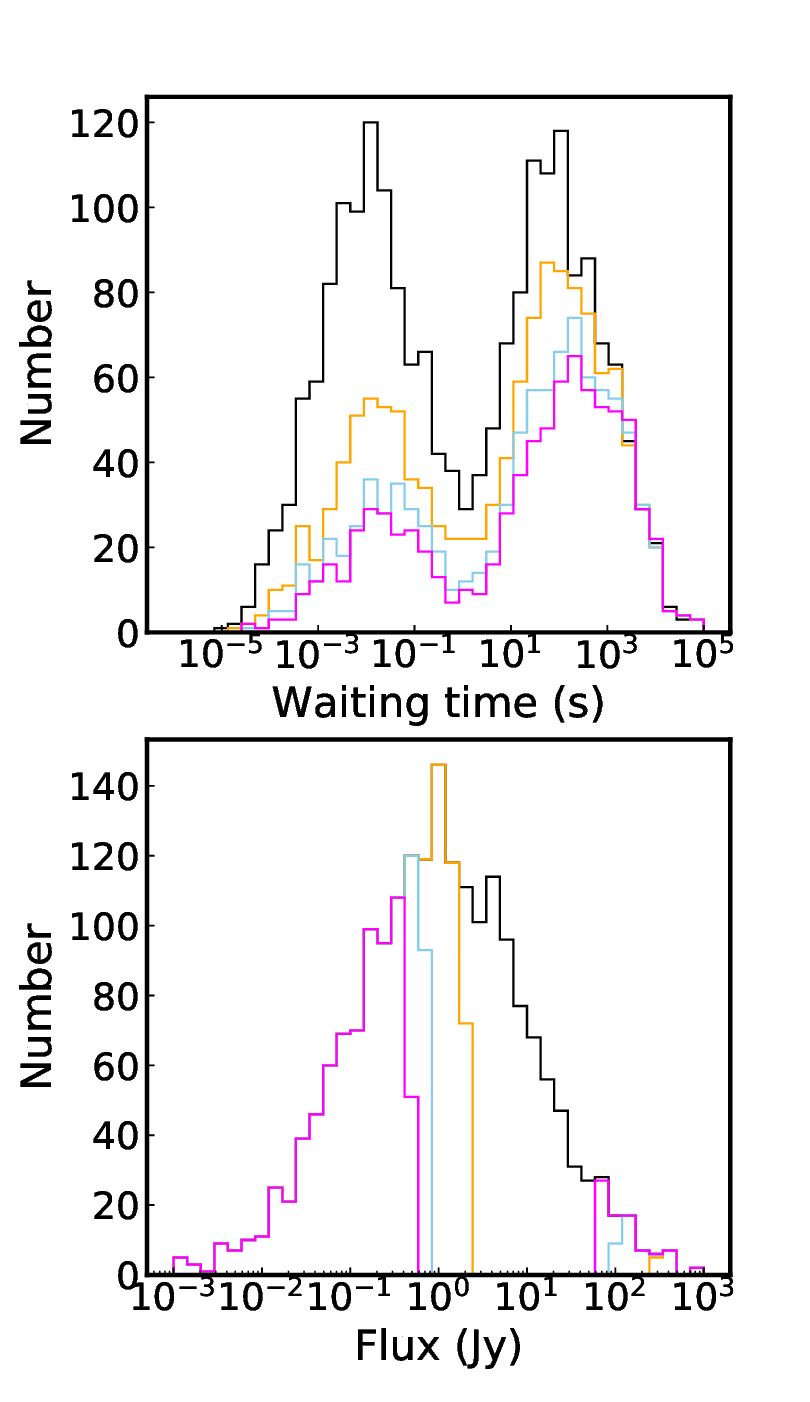}
		\caption{Similar to Figure  \ref{fig1} but fixing $\gamma_{\rm p}=100$ and letting $\theta_B$ vary. The surface dipole field $B_{\rm s}=5\times 10^{13}\,\rm G$ is adopted. The orange, sky blue and magenta colors represent for $\theta_B=0.05,\,0.08,\,0.1$ cases respectively. The absorption happens in an intervening flux range. The initial distributions is in black color.}
	\end{center}
\end{figure}

We note that the scattering optical depth is not a monotonous function of FRB flux. For a very luminous burst, sometimes $R_\theta$ can exceed $R_{\rm LC}$, therefore within the whole magnetosphere the scattering is negligible due to very small cross section then it can also escape. In Figure \ref{fig5} we mock an another burst sample of same size as in Figure \ref{fig1}, but with a different flux distribution containing more luminous bursts characterized by $\log (F_{\nu,c}/{\rm Jy})=0.0$ and $\log(\Delta t_{1,c}/{\rm s})=-2$. We adopt parameters of $B_{\rm s}=5\times10^{13}\,\rm G$, $f=10^{-3}$ and plot three cases of $\theta_B=0.05,\, 0.08,\, 0.1$ while fixing $\gamma_{\rm p}=100$. All other parameters are adopted the same as in Figure \ref{fig1}. We can clearly see that some very luminous bursts could escape the magnetosphere under suitable conditions. Same as Figure \ref{fig1}, faint bursts can escape freely and the left peak of waiting-time distribution drops much more than the right peak.

\section{Comparison with Observations}
\label{sec4}
\begin{figure*}
	\label{fig6} \centering\includegraphics[angle=0, width= 1.0\textwidth]{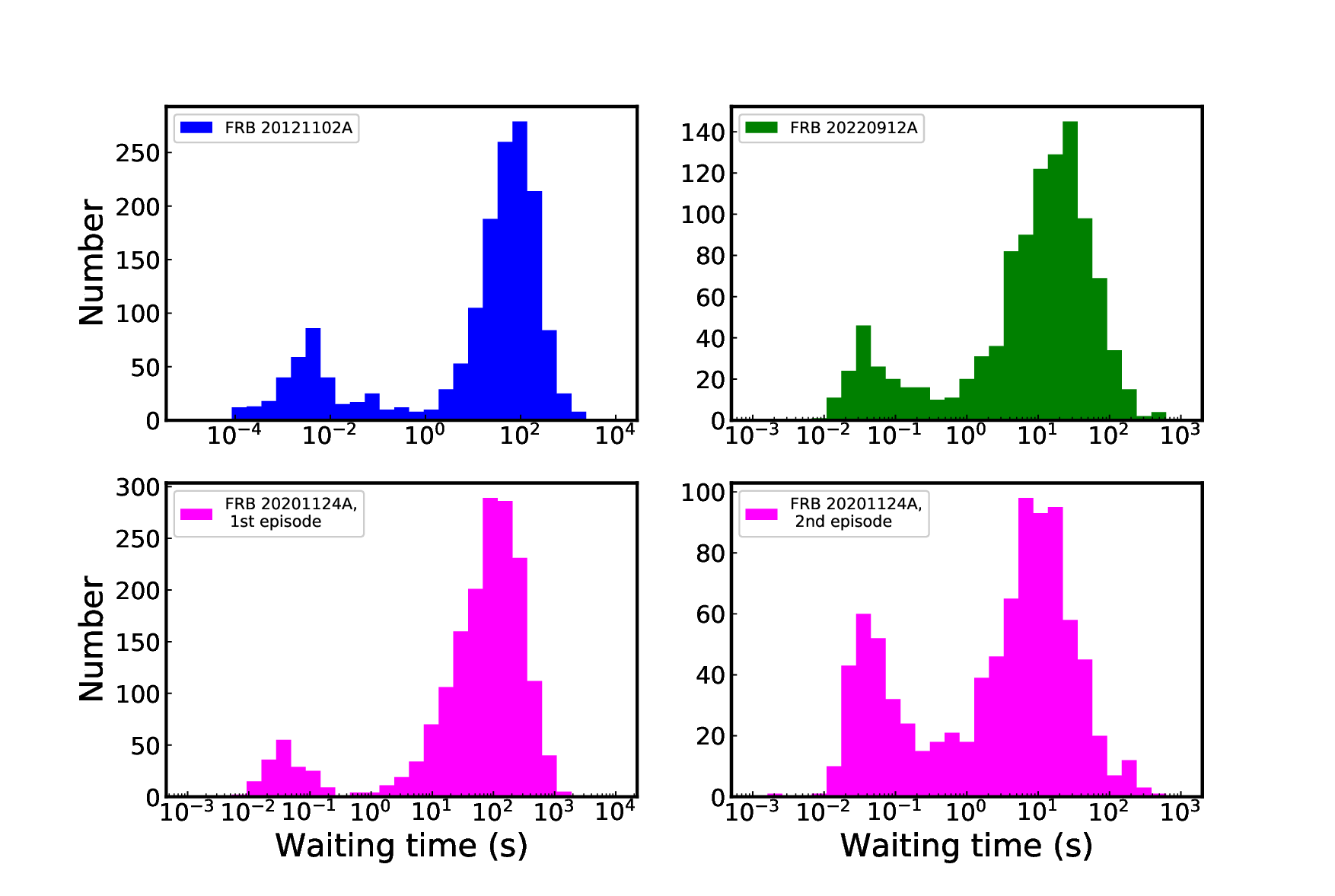}
	\caption{The observed waiting-time distributions of four FAST samples, all of which show a universal left-low-right-high double-peaked character.}
\end{figure*}

\begin{figure*}
	\label{fig7} \centering\includegraphics[angle=0, width= 1.0\textwidth]{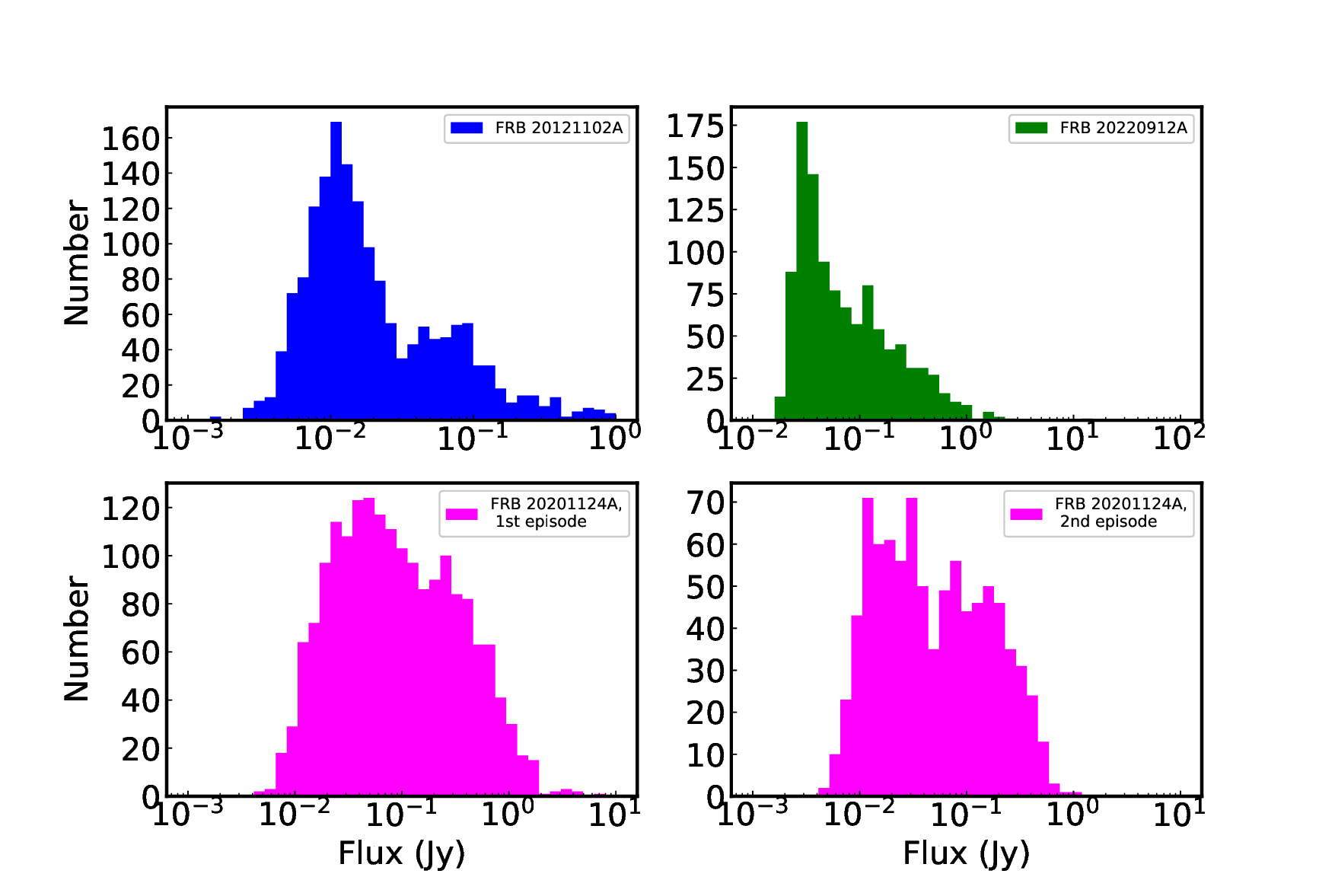}
	\caption{The observed flux distributions of four FAST samples, neither of which can be described by a simple function. However, all of them show ``dip" features at certain fluxes, which might be the imprint of moderate absorption.}
\end{figure*}

\begin{figure}
\label{fig8}
\begin{center}
\includegraphics[width=0.45\textwidth]{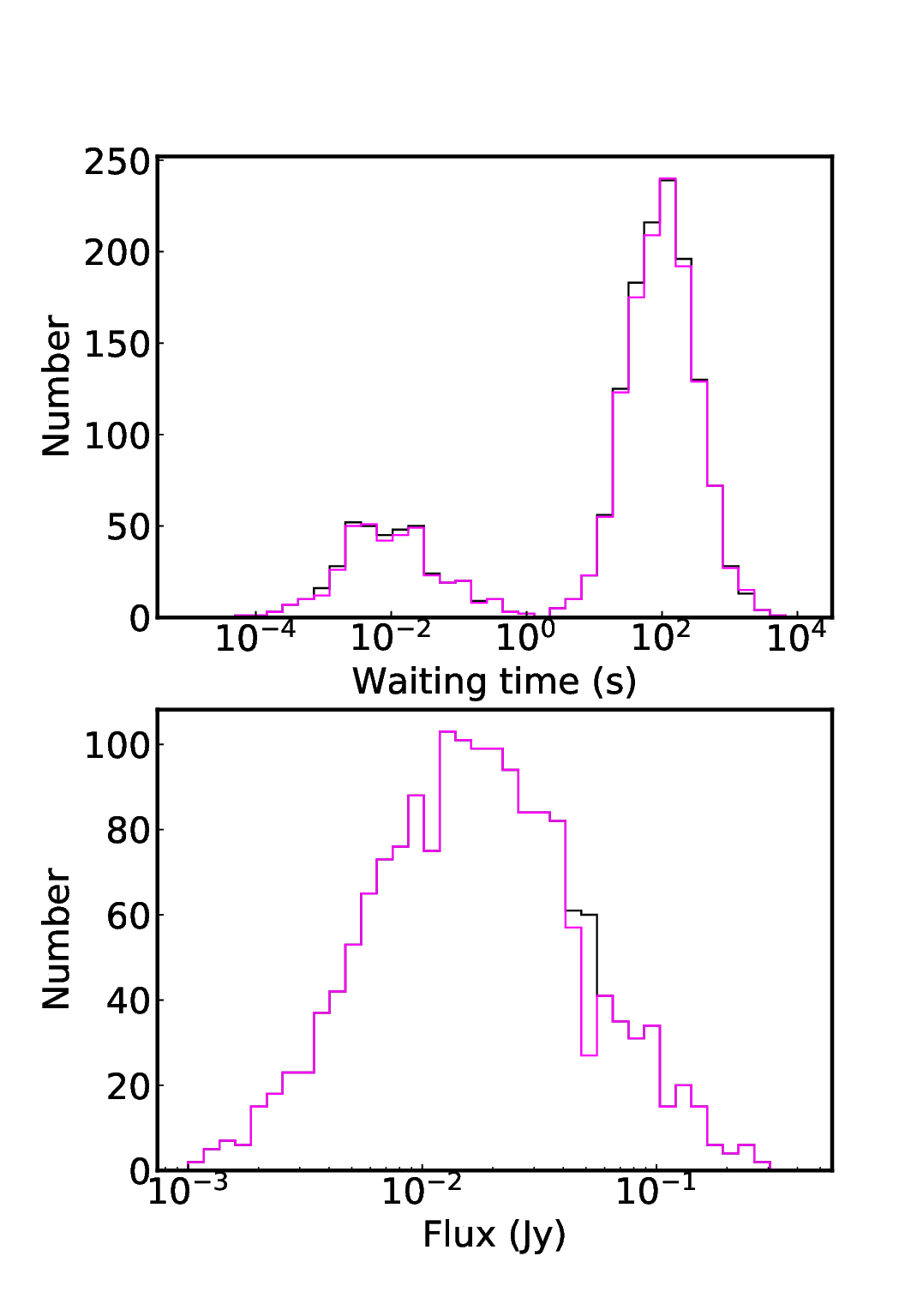}
\caption{An imitation of the distributions resembling FRB 20121102A. Adopting the parameters of $B_{\rm s}=8\times10^{12}\,\rm G$, $f=10^{-5}$, $\theta_B=0.5$, $\gamma_{\rm p}=2300$, the absorption is only moderate and a ``dip" feature in flux distribution is created. }
\end{center}
\end{figure}

FAST has monitered a few active repeaters and the number of bursts from each repeater reached $\sim$ thousands level. These extremely active repeaters are ideal targets for testing the theory. In this work we use the publicly available datasets for three repeating FRB 20121102A \citep{LiD2021}, 20201124A \citep{XuH2022, ZhangYK2022} and 20220912A \citep{ZhangYK2023}. FRB 20201124A has two active episodes and we study these two samples separately. In Figure \ref{fig6} and \ref{fig7} we plot the waiting-time and flux distributions of above four burst samples observed by FAST. All four samples show double-peaked waiting-time distributions with left lower than right, which is quite consistent with our model predictions. Also, all four flux distributions deviate from a simple log-normal shape and there seems moderately-absorbed ``dip" features. Here we just show an example to simulate the dip feature resembling the burst sample of FRB 20121102A from \citet{LiD2021}. We mock a random sample of size 1700 with log-normal flux distribution $\log (F_{\nu,c}/{\rm Jy})=-1.8$, $\sigma_F=0.45$ and bimodal waiting-time distribution $\log(\Delta t_{1,c}/{\rm s})=-2$, $\sigma_{t,1}=0.75$, $\log(\Delta t_{2,c}/{\rm s})=2$, $\sigma_{t,2}=0.5$. Then we show a case of $B_{\rm s}=8\times10^{12}\,\rm G$, $f=10^{-5}$, $\theta_B=0.5$, $\gamma_{\rm p}=2300$ and substituting $D_{\rm L}=0.949\,\rm Gpc$ for FRB 20121102A, the initial and absorbed distributions are represented by black and magenta colors respectively in Figure \ref{fig8}. Similar dip feature appears in the flux distribution and the absorption is just moderate. However, it is difficult to deduce the initial FRB distributions from the observed ones and make any reliable constraint on physical parameters because they are highly degenerated and uncertain. All other three samples show similar dips that may be caused by absorption, and recently the evidence of this feature has been strengthened by the observation of FRB 20201124A by other telescopes \citep{Kirsten2023}. 

\section{Discussion and Conclusions}
\label{sec5}
In this work we revisit the propagation of FRBs in the magnetosphere and show that the induced magnetic field can play an important role. This field is caused by the circular motion of plasma electrons under ponderomotive force and will gradually decay as the electric current disspates due to finite resistivity. The exsitence of $B_{\rm ind}$ will reduce the effective field strength and enlarge the possibility of scattering absorption, and this effect is of particular importance for consecutive bursts with time-lags shorter than milliseconds. We find that the possiblity of FRBs being unscattered in the magnetosphere under typical physical condition of a magnetar is just acceptable, however to be more specific, neither as pessimistic as \citet{Beloborodov2021} nor as optimistic as \citet{Qu2022}. 

This effect of an induced magnetic field leads to a preferential absorption on bursts with relatively high luminosity and very short time-lag. Therefore, whether a burst can escape depend both on the property of itself and previous bursts. Observations of active repeaters indicate bimodal waiting-time distributions with lower left peaks, which is quite consitent with our prediction. However, we have not made any conclusions on what the initial waiting-time distributions are. The initial double log-normal function in this work is just a tentative choice for two stochastic FRB production processes. We note that other initial distributions can also be modified by absorption. Further, the small ``dip" structures on the flux distributions of these repeaters may be also in favor of absorption. 

FRBs may go through many other kinds of absorption processes in the magnetosphere \citep{Lyutikov2021c}, and also on their path to Earth such as free-free and synchrotron-self absoption. As long as absorption occurs, it will leave a nonnegligible imprint on the waiting-time distribution. Therefore these processes may also contribute to shape the observed distribution. What makes the absorption discussed in this paper unique is that the adjacent bursts are no longer independent events and the former burst will influence the absorption possiblility of the latter one. The shorter the time-lag, the stronger the influence will be. 

However, one may still argue that the observed distributions are just the intrinsic ones at the FRB production site and no absorption is needed. However, all four FAST samples show similar bimodal waiting-time distributions is unusual, as also confirmed by Arecibo observation \citep{Hewitt2022}. A smoking-gun evidence of our model is that in the future a double-peaked flux distribution as in Figure \ref{fig5} is once observed for an active repeating FRB.

\acknowledgements

We acknowledge Bing Zhang for constructive discussion and an anonymous referee for helpful comments. This work is supported by National SKA Program of China (2022SKA0130100), the National Natural Science Foundation of China (Grant Nos. 12373052, 12321003, 12041306, 11833003), the CAS Project for Young Scientists in Basic Research (Grant No. YSBR-063).

\bibliographystyle{aasjournal}
\bibliography{FRBlatest.bib}

\end{document}